\documentclass[a4paper]{article}
\usepackage[ansinew]{inputenc}
\usepackage{times}
\usepackage[T1]{fontenc}
\usepackage{graphicx}
\usepackage{geometry}
\usepackage{amssymb}
\usepackage{amsmath}

\usepackage{rotating}

\usepackage{xcolor}
\colorlet{shadecolor}{gray!25}
\usepackage{subfig}
\usepackage{float}
\usepackage{lscape}

\author{Ludwig A. Hothorn,\\ 
Im Grund 12, D-31867 Lauenau, Germany\\ \scriptsize(retired from Leibniz University Hannover)}

\title{Closed test procedures for the comparison of dose groups against a negative control group or placebo}
\begin{document}

\maketitle
\begin{abstract}
Dose groups are compared with a control assuming an order restriction usually by the Williams
trend test. Here, as an alternative, two variants of the closed testing procedure are considered, one where global Williams tests are used in the partition hypotheses, and another where pairwise contrast tests are used for this purpose. Related R software is provided.
\end{abstract}

%%%%%%%%%%%%%%%%%%%%%%%%%%%%%%%%%%%%%%%%%%%%%%%%%%%%%%%%%%%%%%%%%%%%%%%%%%%%%%%%%%%%%%%%%%%%%%%%%%%%%%
\section{The problem}\label{sec1}
Comparisons of $k$ dose groups with a negative control assuming a monotonic dose-response relationship are often performed in biomedical experiments by means of the Williams trend test \cite{Williams1971}. For example, organ weights of rats in three concentrations of pentabromodiphenyl ethers in a toxicological bioassay \cite{Dunnick2018} or the reduction of serum uric acid in gout-free subjects considering two doses of tuna extract compared to placebo \cite{Kubomura2016} in a randomized clinical trial. 
One uses the Williams test instead of the Dunnett test \cite{Dunnett1955} (without the assumption of an order restriction) because of the higher power (due to the restriction of $H_1$) and especially the possibility to interpret a trend (both globally and locally for selected parts of the dose-response relationship). The main difference between the Dunnett test and the Williams test is that the former considers comparisons between $C$ and individual $D_i$, but the latter does not consider comparisons with explicit doses, only pooled doses (except $D_{max}-C$). Therefore, an order-restricted test is derived here for the comparison to the control with the individual doses.
The closed test procedure \cite{MARCUS1976} (CTP) is an alternative to the max-T test for multiple contrasts on which the Williams test is based \cite{Bretz2006}. Two special cases of CTP are considered here: the complete hypothesis family when comparing to control alone \cite{Sonnemann2008} and the decision tree reduction when assuming an order restriction \cite{Hothorn1997}. Thus, related closed test versions of order-restricted tests are derived here. 

%%%%%%%%%%%%%%%%%%%%%%%%%%%%%%%%%%%%%%%%%%%%%%%%%%%%%%%%%%%%%%%%%%%%%%%%%%%%%%%%%%%%%%%%%%%
\section{A brief description of the Williams procedure}
In the original paper \cite{Williams1971} the procedure was described for maximum likelihood estimators under order restriction, but one can formulate this simplified as multiple contrast test (MCT) \cite{SHAFFER1977}, \cite{Mukerjee1987}. 
The basis is a maximum test over several t-distributed standardized contrast tests $t_q$ (here referred to as maxT-test): $t_{MCT}=max(t_1,...,t_{q'})$ with  $	t_q=\sum_{i=0}^k c_i\bar{y}_i/S \sqrt{\sum_i^k c_i^2/n_i}$ where $c_i^q$ are the contrast coefficients (see selected examples below). A adjusted p-values are given by the minimum empirical $\alpha$-level: $\frac{\sum_{i=0}^k c_i\bar{y}_i}{S\sqrt{\sum_i^k c_i^2/n_i}} = t_{q,df,R,1-sided,1-min(\alpha)}$ where $t_{q,df,R,1-sided,1-\alpha}$ is the quantile of central q-variate t-distribution, easily available in the package mvtnorm \cite{Mi2009}. Compatible to the adjusted p-values should be simultaneous confidence intervals. They are not considered here because of their difficulties in general closed tests \cite{Guilbaud2018}.\\

%%%%%%%%%%%%%%%%%%%%%%%%%%%%%%%%%%%%%%%%%%%%%%%%%%%%%%%%%%%%%%%%%%%%%%%%%%%%%%%%%%%%%%%%%%%%
\section{A brief description of restricted closed testing procedures}
Starting point of any CTP is the a problem-adequate definition of the interesting elementary hypotheses, here: $H_i: \mu_i-\mu_0$.
In a second step a decision tree containing all subset intersection hypotheses up to the global hypothesis, involving these elementary hypotheses is constructed \cite{MARCUS1976}. $H_i$ is rejected  at level $\alpha$ if and only if $H_i$ itself is rejected and all hypotheses which include them in the decision tree (again each at level  $\alpha$). Each hypothesis is tested with a level $\alpha$-test, with any appropriate test - this allows a high flexibility of the here described approach. Each of these tests (determined by the $\xi$ elementary hypotheses) represents an intersection-union test (IUT), i.e. $T^{CTP}=min(T_1,...,T_{\xi})$, or more common $p^{CTP}=max(p_1,...,p_{\xi})$. Here this approach is demonstrated for a rather simple design with $k=2$, the family include the following elementary (e.g. $H_0^{01}$), intersection (e.g. $H_0^{012}$) and global hypotheses (e.g. $H_0^{0123}$): \\
$H_0^{01}: \mu_0=\mu_1 \subset [H_0^{012}, H_0^{013}] \subset H_0^{0123}$\\
$H_0^{02}: \mu_0=\mu_2 \subset [H_0^{012}, H_0^{023}] \subset H_0^{0123}$\\
$H_0^{03}: \mu_0=\mu_3 \subset [H_0^{013}, H_0^{023}] \subset H_0^{0123}$\\

Monotonic order constraint $H_1: \mu_0\leq \mu_1\leq...\leq\mu_k|\mu_0<\mu_k$ (for any possible pattern of equalities/inequalities) further greatly simplifies this specific CTP. Under this restriction of $H_1$, rejection of $H_0^{0123}$ implies rejection of $H_0^{013}$ and $H_0^{03}$, and rejection of $H_0^{012}$ implies rejection of $H_0^{02}$, and so on. The hypothesis system is highly simplified:\\
$H_0^{01}: H_0^{01} \wedge H_0^{012} \wedge H_0^{0123}$\\
$H_0^{02}: H_0^{012} \wedge H_0^{0123} $\\
$H_0^{03}: H_0^{0123}$\\

For these hypotheses, \textit{any} one-sided level $\alpha$ test can be used. The elementary hypotheses should be tested with contrast tests for $\mu_i-\mu_0$, not with two-sample tests, so as not to reduce power in small $n_i$ designs. For the partition and global hypothesis, any order-constrained test can be used. Because of the $D_i-C$ comparison, two versions of the special CTP are considered here: i) Williams global test for each subset (denoted CW), ii) pairwise contrast tests for $\mu_{\xi}-\mu_0$ (where $\xi$ is the highest dose in each subset) (denoted CP).

%%%%%%%%%%%%%%%%%%%%%%%%%%%%%%%%%%%%%%%%%%%%%%%%%%%%%%%%%%%%%%%%%%
\section{Simulation study}
Empirical power and size of these tests are demonstrated by a tiny simulation study for a low-dimensional one-way design $y_{ij}=\mu+ \text{factor}_i+\epsilon_{ij}$ ($i=0,...,k$) with $\epsilon_{ij} \propto N(\mu_i, \sigma^2)$. Random experiments with a single primary endpoint $y_{ij}, k=2$ in a balanced design were used, estimating their per-pair power $\pi_{01}, \pi_{02},\pi_{03}$ for six strictly monotonic alternatives and two shapes with a downturn effect at the high dose (occuring in some in-vitro toxicity assays).
Common simulation studies in the framework of simultaneous inference compare any-pair power \cite{Hothorn2020} or average power \cite{Stevens2017}. These concepts greatly simplify the power comparisons, but are not purposeful because they do not take into account which individual comparison is currently in the alternative. However, one does not want to know whether any dose is different from the negative control. No, one wants to evaluate exactly a particular dose relative to the control. That is why the concept of per-pair power is used here, although it is difficult to interpret (and that is why k=2 was used).\\
The four tests are abbreviated as D (Dunnett original), W3 (Williams $D_3-C$ contrast only, since the other contrasts are non-pairwise), CW (CTP using a subset of Williams global tests), and CP (CTP using pairwise contrasts), where $D_i$ is pairwise power ($D^a$ is any-pairs power for the Dunnett test for reference purpose). Instead of complete power curves, only one relevant point in the alternative is considered for about $\pi>0.8$.

\begin{table}[ht]
\centering\scriptsize
\begin{tabular}{l|l|rrr|r|rrr|rrr}
  \hline
Shape& $H_1$ & $D_1$ & $D_2$ & $D_3$ &  $W_3$ & $CW_1$ & $CW_2$ & $CW_3$ & $CP_1$ & $CP_2$ & $CP_3$  \\ 
  \hline
$H_0$&$\mu_0=\mu_1=\mu_2=\mu_3$&0.02 &0.02 & 0.03 &0.03 & 0.01 & 0.02& 0.05 &0.01  & 0.05 & 0.05 \\ 
\hline\hline
	
Monot &$\mu_0<\mu_1=\delta\ <\mu_2=\mu_3=3\delta$ 	& 0.10 & 0.82 & 0.81  & 0.85 & 0.18 & 0.86 & 0.95 & 0.18 & 0.84 & 0.90 \\ 
&$\mu_0<\mu_1=2\delta\ <\mu_2=\mu_3=3\delta$						& 0.42 & 0.80 & 0.81  & 0.85 & 0.57 & 0.87 & 0.95 & 0.54 & 0.83 & 0.91  \\ 
&  $\mu_0=\mu_1=\mu_2<\mu_3=3\delta$										& 0.02 & 0.02 & 0.81  & 0.85 & 0.02 & 0.05 & 0.85 & 0.01 & 0.05 & 0.90  \\ 
&  $\mu_0<\mu_1=\mu_2=\mu_3=3\delta$										& 0.81 & 0.82 & 0.82  & 0.86 & 0.88 & 0.94 & 0.96 & 0.80 & 0.85 & 0.91  \\ 
&  $\mu_0=\mu_1<\mu_2=\mu_3=3\delta$										& 0.02 & 0.82 & 0.80  & 0.85 & 0.04 & 0.86 & 0.95 & 0.03 & 0.84 & 0.90  \\ 
 & $\mu_0<\mu_1=\delta <\mu_2=2\delta<\mu_3=3\delta$		& 0.09 & 0.42 & 0.80  & 0.85 & 0.16 & 0.53 & 0.89 & 0.15 & 0.55 & 0.90 \\ 

\hline
Non-m &$\mu_0=\mu_1<\mu_2=3\delta>\mu_3=2\delta$			& 0.02 & 0.80 & 0.43 &  0.49 & 0.04 & 0.78 & 0.82 & 0.03 & 0.57 & 0.59  \\ 
 &$\mu_0=\mu_1<\mu_2=3\delta>\mu_3=\delta$ 							& 0.02 & 0.81 & 0.10  & 0.14 & 0.03 & 0.62 & 0.62  & 0.02 & 0.19 & 0.19 \\ 
\hline
\end{tabular}
\caption{Per-comparison power estimates for selected alternatives}

\end{table}

By definition, all tests control the familywise error rate (not shown in detail here). For strictly monotonic alternatives, the power of the Williams test is by definition slightly greater than that of the Dunnett test (directly comparable only for $D_3-0$). Both CTP-tests almost always show superiority in power for all $\pi_i$, and for some patterns a marked superiority over Dunnett's test. As expected, the power of the CW test is slightly better than that of the CP test, but the latter being impressive for its simplicity.  Depending on the magnitude of the response decline in $D_{max}$, all tests assuming an order restriction are not robust, as expected.

%%%%%%%%%%%%%%%%%%%%%%%%%%%%%%%%%%%%%%%%%%%%%%%%%%%%%%%%%%%%%%%%%%%%%%%%%
\section{Evaluation of a data example}
Relative liver weights in male rats of 4 dose groups and a negative control (abbreviated as 1) of an in vivo bioassay are used as a data example (data available in library(nparcomp)). The boxplots in Figure 1 show an approximately symmetrical distribution and homogeneous variances, so the standard tests are used.

\begin{figure}[htbp]
	\centering
		\includegraphics[width=0.40\textwidth]{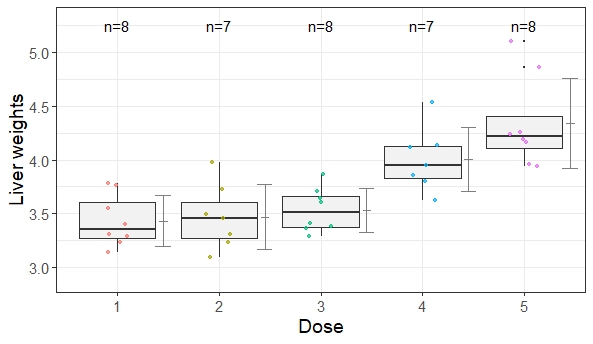}
	\caption{Jittered boxplots for liver weights}
	\label{fig:liverBOX}
\end{figure}

\begin{table}[ht]
\centering\small
\begin{tabular}{r|llll}
 Comparison& Dunnett  & Williams  & CW  &CP  \\ \hline
 5-1& $ 1.99^{-6}$& $ 8.8^{-7}$ & $ 8.8^{-7}$ & $ 4.7^{-7}$ \\
 4-1& $ 1.51^{-3}$  & NA & $ 7.1^{-4}$ &$ 4.3^{-4}$ \\
 3-1& 0.531  & NA & 0.322 &0.253\\ 
 2-1& 0.719 & NA & 0.406 & 0.406\\
 \hline
\end{tabular}
\caption{Adjusted p-values for liver weight data \small (NA... not available)}
\end{table}
The one-sided multiplicity adjusted p-values for the 4 elementary hypotheses $\mu_i-\mu_0$ are given in Table 2. As expected, the p-value $p_{5-1}$ is smaller in the Williams test and both closure tests than in the Dunnett test. The p-values for the 4 elementary hypotheses $\mu_i-\mu_0$ are smaller in CP than in CW, not surprising for such a step-wise shape of the dose-response relationship. p-values for the elementary hypotheses $[\mu_4-\mu_1], [\mu_3-\mu_1], [\mu_2-\mu_1]$ are not available per definition for the Williams-type test. The R-code of this example is given in the Appendix.

\normalsize
%%%%%%%%%%%%%%%%%%%%%%%%%%%%%%%%%%%%%%%%%%%%%%%%%%%%%%%%%%%%%%%%%%%%%%
\section{Conclusions}
Obviously, no uniformly powerful test can exist for any pattern of monotonic $H_1$, certainly not for alternatives with downturns at high dose(s). In particular, if one considers the specific pattern of dose-response as a priori unknown, the CTP's proposed here prove to be a powerful alternative. In particular, the availability of adjusted p-values for the elementary hypotheses makes these tests attractive. Further generalizations for generalized linear mixed effect model (glmm) (e.g., for proportions \cite{Hothorn2020k}), use for estimating the no-observed-adverse-event-level (NOAEL), or consideration of trend tests for modeling dose as a quantitative covariate \cite{Schaarschmidt2020}, and a software implementation will follow shortly.

\footnotesize

\bibliographystyle{plain}
\footnotesize
%\bibliography{D:/PUB/Dunnett2020/Dunnett2020}

%\bibliographystyle{plain}
%\begin{thebibliography}{10}

%\bibitem{Bauer1996}
%P.~ Bauer and M.~ Kieser.
%\newblock A unifying approach for confidence intervals and testing of equivalence and difference.
%\newblock {\em Biometrika} 83: 934 -- 937, 1996.
%\end{thebibliography}
\small
\section*{Appendix: R-code for the data example}
\tiny
\begin{verbatim}
library(nparcomp)
library(multcomp)
data(liver)
liver$dose<-as.factor(liver$dosage)
mod1<-lm(weight~dose,data=liver)
CM04 <- c(-1,0,0,0,1)
CM03 <- c(-1,0,0,1,0)
CM02 <- c(-1,0,1,0,0)
CM01 <- c(-1,1,0,0,0)
ni<-aggregate(weight ~ dose, data = liver, length)$total
cmat0123<-contrMat(ni[1:4], type="Williams"); V4 <-c(0,0,0)
Cmat0123<-cbind(cmat0123,V4)
cmat012<-contrMat(ni[1:3], type="Williams"); V3 <-c(0,0)
Cmat012<-cbind(cmat012,V3, V3)

T04<-summary(glht(mod1, linfct = mcp(dose= CM04), alternative="greater"))$test$pvalues
T03<-summary(glht(mod1, linfct = mcp(dose = CM03), alternative="greater"))$test$pvalues
T02<-summary(glht(mod1, linfct = mcp(dose = CM02), alternative="greater"))$test$pvalues
T01<-summary(glht(mod1, linfct = mcp(dose = CM01), alternative="greater"))$test$pvalues
W01234<-min(summary(glht(mod1, linfct = mcp(dose ="Williams"), alternative="greater"))$test$pvalues)
W0123<-min(summary(glht(mod1, linfct = mcp(dose =Cmat0123), alternative="greater"))$test$pvalues)
W012<-min(summary(glht(mod1, linfct = mcp(dose =Cmat012), alternative="greater"))$test$pvalues)
W01<-min(summary(glht(mod1, linfct = mcp(dose = CM01), alternative="greater"))$test$pvalues)
CTP4<-T04
CTP3<-max(T04,T03)
CTP2<-max(T04,T03, T02)
CTP1<-max(T04,T03, T02, T01)

CTW4<-W01234
CTW3<-max(W01234,W0123)
CTW2<-max(W01234, W0123, W012)
CTW1<-max(W01234, W0123, W012, W01)

Du<-summary(glht(mod1, linfct = mcp(dose ="Dunnett"), alternative="greater"))$test$pvalues
Wi<-summary(glht(mod1, linfct = mcp(dose ="Williams"), alternative="greater"))$test$pvalues
\end{verbatim}

\end{document}